# Characterizing Phase Transitions in Liquid Cesium by a Soft-core and Large Attractive Equation of State

(Department of Chemistry Shiraz University, Shiraz 71454, Iran)

M.H. Ghatee*, and M. Bahadori

E-mail : ghatee@sun01.susc.ac.ir
Fax : +98 711 228 6008

---
* To whom all correspondence should be addressed

# Characterizing Phase Transitions in Liquid Cesium by a Soft-core and Large Attractive Equation of State

*Abstract*

*This paper investigates to identify phase transitions in condensed liquid cesium metal by considering the variation of intermolecular potential parameters $\varepsilon$ and $r_m$ in the whole liquid range, with $\varepsilon$ being the potential well-depth and $r_m$ the position of minimum potential. These parameters were obtained from the parameters of a new equation of state that was derived recently by using the characteristic potential function. By this method, transitions at about 575 K, 800 K, 1000 K, 1350 K and 1650 K were identified. Transitions at 575 K, 800 K, and 1000 K are weak but, the one at 1350 K is very significant and has been explored experimentally and theoretically as the metal non-metal transition (MNMT), which is a phase transition before the critical condition dominates the thermodynamics. Also variations of the linear correlation coefficient of the isotherms generate a spot point pattern of these transitions. Our observations at 575 K for $\varepsilon$ and $r_m$ are in accord with the anomalies in adiabatic thermal coefficient of pressure, density, viscosity, electrical conductivity, and structure factor.*

# 1. Introduction

Physical chemists are constantly developing new theoretical methods to unravel the complex behavior of liquid alkali metals, which are composed of many simple interacting atoms. The electronic structure of alkali metals depends strongly on their thermodynamic state and has been the central focus for new theories. Several investigations have been performed to the temperature and density dependence of the nature of interactions in condensed liquid alkali metals [1,2,3]. Cesium and rubidium have been investigated more than the other alkali metals because of their lower critical temperature [2,3].

The most important change in the electronic structure and the nature of interatomic interactions in molten alkali metals is the metal non-metal transition (MNMT) that occurs near the critical region. During a metal non-metal transition uniform electronic structure of metals are disturbed, and polyatomic clusters of different sizes are formed. In these clusters nearly free electron approximation of valence electrons was omitted and electrons become more correlated to their parent atoms thus the system shows a non-metallic behavior.

In addition to MNMT another weaker transition in condensed liquid cesium based on the experimental observations was reported recently. New experimental data on the adiabatic thermal coefficient of pressure of liquid cesium [4,5] has revealed this second order phase transition near 590 K [5]. Also thermal neutron scattering data has been used to establish temperature dependence of the structure factor of liquid cesium [6,7].

Liquid alkali metals are treated statistical mechanically as an ensemble of metal atoms. As the temperature is increased the valance electrons gradually become more correlated with their parent ions, leading to polarizable polyatom clusters. Therefore, a soft repulsive and large attractive potential function should describe the interaction potential of these systems. Exp6 potential function has such properties. Exp6 potential function has been applied to derive the linear Exp6-isotherms [8]. Application of Exp6-isotherm to liquid cesium was successful in the metallic region as well as beyond the onset of MNMT region. This feature has been attributed to the well estimated attraction which is increased as the result of increase in the number of polarized metal atoms as the liquid metal is expanded.

As a part of continuing studies, recently we have also proposed a potential function in agreement to the characteristics large attraction and soft repulsion at the asymptotes of interactions, and then accordingly have derived a new equation of state [9]. This equation of state is superior to the one that uses Exp6 potential function. Because of its better estimation of attractive interaction it applies remarkably in MNMT region and at the critical temperature $T_c$. Parameters of the derived equation of state are temperature dependent and we have used them to obtain the molecular parameters of the potential function. Magnitudes of evaluated molecular parameters are functions of temperature and their plots can be used to identify transitions revealing the region of MNMT in condensed liquid cesium.

In the present paper we apply our method to unravel phase transitions in condensed liquid cesium in the temperature range 350-1950 K. These transitions comprise a MNMT at about 1350 K. The method is still capable of recognizing a second order phase transition between 550 K and 600 K, which has recently reported in reference 5. Furthermore, the linear correlation coefficients of the isotherms of liquid cesium system over the whole liquid range indicate the occurrence of these transitions.

**2. The Potential Function and Equation of State**

An effective potential function for expanded liquid cesium has been obtained from structure factor by the inversion method [1]. These data suggest density dependence of the effective interactions and the related analysis has indicated that at large interatomic distance the potential function oscillates and takes $(1/r^3)$ asymptotic form. On the other hand, by integral equation theory the effective pair potential of liquid cesium has been derived over a wide range of temperature from experimental structure factors by inverse method [10]. The repulsive side of this potential has been analyzed in term of ($a/r^m$), where *a* is a constant and m=7.7, 6.8, 5.6 at 773 K, 1373 K and 1673 K, respectively.

From these experimental data and from the fact that the range of liquid cesium densities correspond to interatomic distances around $r_m$, we have proposed the potential function Eq. (1) [9]:

$$u(r) = 4\varepsilon \left[ \left(\frac{\sigma}{r}\right)^6 - \left(\frac{\sigma}{r}\right)^3 \right]. \qquad (1)$$

In Eq. (1) $\sigma = 2^{-1/3} r_m$, $r_m$ is the interatomic distance at which potential is a minimum, and $\varepsilon$ is the potential well depth. This two-parameter potential function has shown to describe appropriately the thermodynamic properties of liquid cesium in the whole liquid range, including MNMT region and at the critical temperature [9].

We have calculated the total interaction potential energy U, of the liquid state comprising N particles by the model in which **1)** the interaction can be approximated by sum of the interactions in pair, and **2)** only the interactions between the nearest neighbors are the most effective interactions. Therefore, $U = \sum_{i>j=1} u(r_{ij})$, where $r_{ij}$ is the distance between atom i and atom j. Performing the sum, it leads to $U = (N/2) u(r)$, where u(r) is the pair interaction potential energy. In this way $\varepsilon$ becomes the binding energy of a central atom interacting to 8 other atoms located at the corner of the cube (of the unit cell).

In this study, by the arguments given in previous paragraphs, the pair potential function defined in Eq. (1) is applied to investigate the properties of liquid cesium metal. It turns out that Eq. (1) becomes an effective pair potential function. Since the coordination number is a function of temperature (and density as well) we will apply the exact coordination number.

The pressure P of the liquid system is calculated by solving the thermodynamic equation of state $(\partial E/\partial V)_T = T(\partial P/\partial T)_V - P$, where V is molar volume, T is the absolute temapatature, and E is the total energy including total interatomic potential energy and the kinetic energy. The kinetic energy is independent of volume, and thus application of the thermodynamic equation of state is simplified. We have derived the equation of state in the final form [9],

$$(Z-1)V^2 = C + B\left(\frac{1}{\rho}\right) \qquad (2)$$

where Z is the compressibility factor, $\rho$ is the molar density. The parameters of the linear isotherm B and C, are temperature dependent and are related to the attraction and repulsion parts of the potential function Eq. (1), respectively. The molecular parameters were calculated using the numerical values of the slope B and the intercept C of the isotherm of the equation of state:

$$r_m = 2^{1/3}\left(\frac{-K_{bcc}^3}{2}\frac{C}{B}\right)^{1/3}, \qquad \varepsilon = \frac{RT}{N}\frac{B^2}{C} \qquad (3)$$

where $K_{bcc}$ is a constant characteristic of the (body center cubic) unit cell of cesium and is equal to $\left(3\sqrt{3}/4N\right)^{1/3}$, with N being Avogadro's number.

In this model, due to the use of experimental PVT data for the calculation of the parameters of linear isotherm Eq. (2) [9], $\varepsilon$ becomes the binding energy of an atom in ensemble of N-1 other similar atoms. Also $\varepsilon$ is a function of T, B, and C, thus a fair balance between interatomic attraction and repulsion is responsible for variation of $\varepsilon$ with temperature. The variation of $r_m$ with temperature is almost linear except some wiggling close to $T_c$. This observation is in accord with the neutron scattering results [2].

Recently this method has been used to derive an equation of state for cesium by using Lennard-Jones (8.5-4) potential function which has been determined by application of cohesive energy density data of liquid cesium including data at the proximity of absolute zero [11,12]. This potential function has been successful in prediction of transport properties of cesium vapor as well as the prediction of equilibrium thermodynamic properties of liquid cesium [12]. Furthermore, this method has been successful in determination of density dependent equations of state for divalent mercury liquid [13].

## 3. Results and Discussion

It has been known that the intermolecular interaction potential functions change with temperature and density. The structural changes are usually taken into consideration by studying pair correlation function. The influence of density and temperature on the intermolecular interactions of normal liquid is inappreciable but is effective on liquid metals. On the other hand, for these metals the nature of interatomic interactions will change as the critical point is approached. The potential energy profile will change as the thermodynamic state approaches to the state of the expanded liquid in the critical region. Accordingly it is expected that the pair interaction potential of the N-body system is influenced more or less in the same way as the total interaction potential energy does by the change in the nature of intermolecular forces.

The functional form of $B$ and $C$ can not be determined by the present method. However, both are linear function of $(1/T)$. $B$ and $C$ are indication of interatomic attraction and repulsion, respectively. In this manner the effective pair potential function Eq. 1 becomes temperature dependent (as well as density dependent). This has become evident by application of experimental PVT data to the Eq. (2). These results are in accord with neutron scattering data used to obtain effective pair interaction potential function [2,10].

We have used the molecular potential parameters $\varepsilon$ and $r_m$, obtained by fitting equation of state [Eq. (2)] to PVT data, to scrutinize transitions in the electronic structure and the nature of interactions or in the atomic configuration in liquid cesium. Transitions are confirmed by the fact that both $\varepsilon$ and $r_m$ as well as the linear correlation coefficient of isotherms [of Eq. (2)] show anomalous behavior in the same temperature region.

Figure 1 shows $\varepsilon$ as a function of temperature determined by using Eq. (3) in combined with Eq. (2). In general, $\varepsilon$ decreases with temperature smoothly. However, it will be shown that there are sharp decreases at about 575 K, 800 K, 1000 K, 1350 K and 1650 K indicating second order transitions. (See discussions on the Figures 2 and 3.) The low temperature transitions look insignificant, however, they are coincide at the same locations where phase transition have been reported [5]. The transitions at high temperatures address a more significant phase transition, which is coincided with the region of metal non-metal transition.

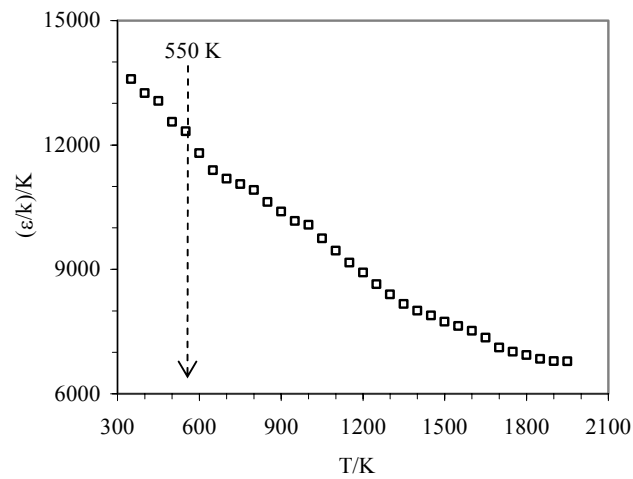

**Figure 1.** Plot of $\varepsilon/k$ versus temperature for liquid cesium [Eq. (3)].

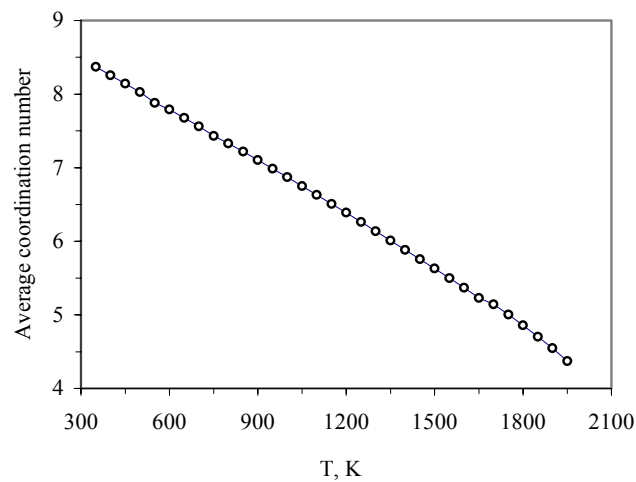

**Figure 2.** Plot of average coordination number versus temperature for liquid cesium (see text for details).

From spectroscopic measurements of the vapor state of cesium and *ab initio* calculations, the value of $\varepsilon/k$ for an isolated pair atoms is determined to be equal to 1589.5 K [14], for which contributions of singlet and triplet type diatomic interactions and the related statistical weighting factor have been regarded.

The magnitude of $\varepsilon$, obtained directly in this study from Eq. (3), is the total interaction energy of an atom with its all-neighboring atoms. If we consider these values as the binding energy of a given cesium atom (e.g., $\varepsilon/k$ =1623 K at T=350 K) within the liquid then it leads almost to the same values for $\varepsilon/k$ (within 2%) reported by the spectroscopic method [14].

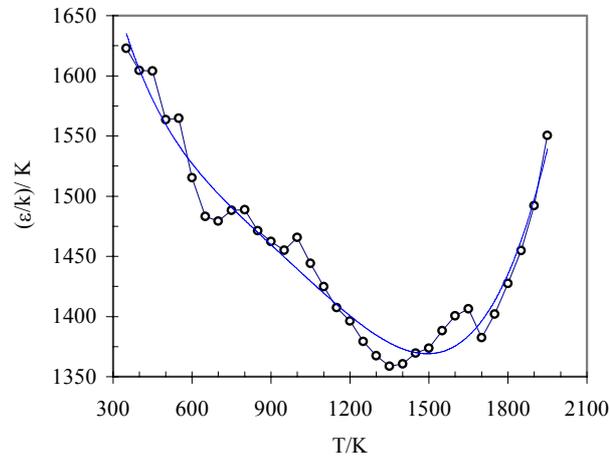

**Figure 3.** Plot of $\varepsilon/k$ scaled by average coordination number versus temperature for liquid cesium. Dot-line this work; doted line follows the general trend of transitions.

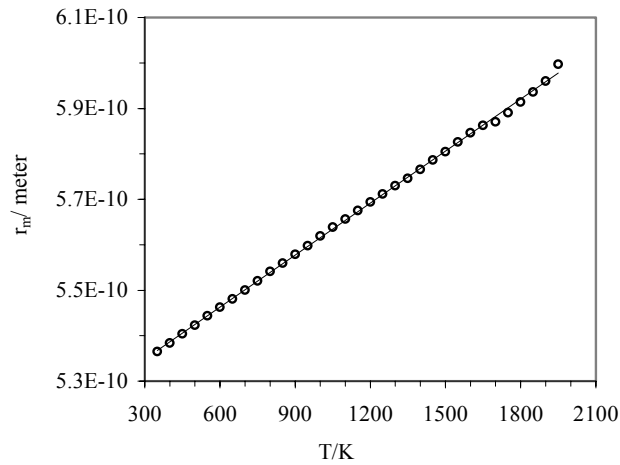

**Figure 4.** Plot of $r_m$ versus temperature for liquid cesium. The solid line is the linear fit through $r_m$ values [Eq. (3)].

Therefore, to have a meaningful characterization of the transition by application of $\varepsilon$, it is essential to have an accurate estimate of the experimental coordination numbers in the whole liquid range of cesium.

Coordination number in the statistical mechanical sense is determined by the integration of experimental pair correlation function up to the principle peak: $C.N. = 2\rho \int_0^{r_m} g(r) 4\pi r^2 dr$, where C.N. is the coordination number and $g(r)$ is the pair correlation function. Then one is certain that the C.N. is the number of nearest neighboring atoms around a central atom. Fortunately from the analysis of the neutron scattering results [15,16], the coordination number of liquid cesium was noted to be a linear function of density [17,18]:

$$C.N. = 0.0043\rho + 0.3478, \qquad \rho \text{ in } (Kg/m^3) \tag{4}$$

Hence by using this equation, at a given temperature, we consider the average of coordination numbers at all available densities in the pressure range 50-600 bar [19] as the coordination number of liquid cesium at that temperature. Such a treatment still leads to a rather linear coordination number with temperature, a plot of which is shown in Figure 2. Then this enables us to calculate the effective potential well depth of the pair potential function Eq. (1) from binding energies, a plot of which versus temperature is shown in Figure 3. By applying coordination numbers (described above), now we see a distinct trend, compared to Figure 1, indicating several transitions, which will be explored further.

Now by turning to the other molecular parameter potential $r_m$ [See Eq. (3).], we see that it is also a function of temperature and varies with temperature linearly as is shown in Figure 4. However, to identify the precise behavior of $r_m$ in the whole range of temperature, the deviations of the evaluated data from a linear fit (see Figure 4) has been estimated and plotted in Figure 5. This Figure indicates that the trend of $r_m$ deviates from a linear behavior with some fluctuations at about 575 K, 800 K, 1000 K, 1350 K and 1650 K, almost coinciding at the same temperatures at which transitions are indicated by the use of the other molecular potential parameter $\varepsilon$.

The principle pattern of $r_m$ with temperature for cesium is linear. It is interesting to note that this is the case for other simple liquid systems, which their $r_m$ values have seen to vary with temperature smoothly. Typically for Ar it is nearly linear, however, in the case of $N_2$ it curves smoothly [20]. These smooth behaviors can be attributed to the smooth changes that occur in the structure and the geometry with the change in density. But for cesium, some considerable wiggling in this pattern are the results of abrupt changes in the electronic structure due to the change in the nature of interactions. The trend of $\varepsilon$, representing an energy balance between attraction and repulsion branches of the potential function, also confirms such transitions.

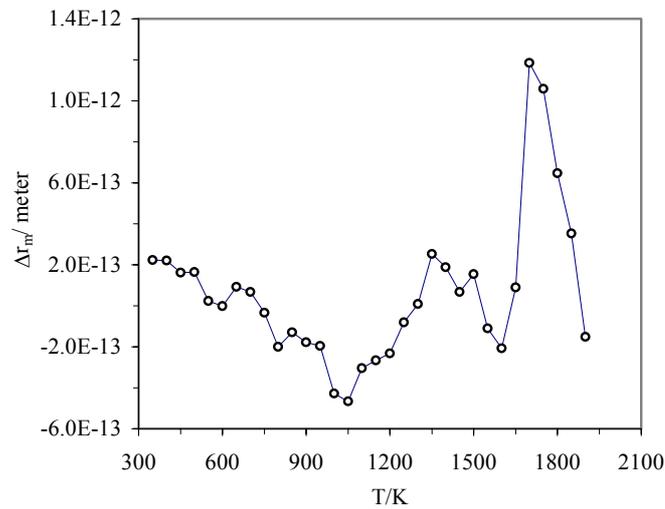

**Figure 5**. Plot of deviations of $r_m$ from the linear behavior versus temperature for liquid cesium.

The transitions indicated by $\varepsilon$ are in agreement alongside with those indicated by $r_m$. The amplitude of the transitions that occur at about 575 K, 800 K, and 1000 K is small and therefore they are weak transitions and their experimental determination need special techniques. Recently, based on adiabatic thermal coefficient of pressure of liquid cesium [4], a transition at the 590 K was reported. A later improvement in their instrumentation has confirmed a second order phase transition clearly [5]. Contrary to experimental difficulties, the present method is capable to estimate pretty well the position of this transition between 550 K and 600 K. The precise position of the transition could be estimated if more experimental PVT data was available. We have furnished a better estimate of the transition temperature by interpolating the PVT data. The trend of the deviation of data from linearity in this region is shown in Figure 6 from which the transition temperature can be determined at about 575 K.

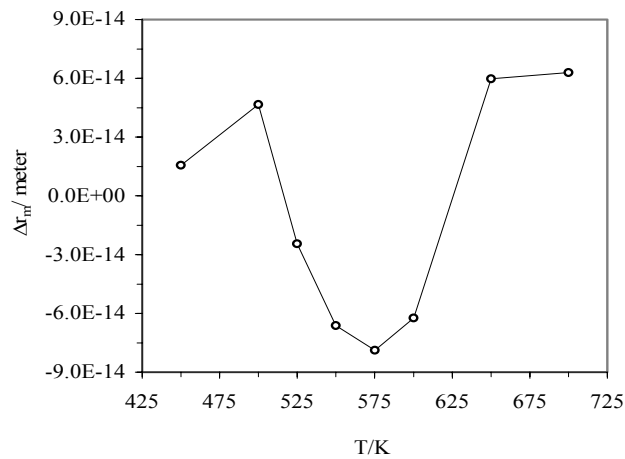

**Figure 6**. The same as figure 5 but for the temperature range 425-725 K.

The amplitude of the transitions at 1350 K and 1650 K is larger and thus they are indication of strong transitions that has been recognized easily [8,9,21-24]. Studies of the electrical conductivity have shown that a continuous MNMT occurs on passing through the critical region. Because of transition at 1350 K, interparticle forces are strongly affected by the change in density. Therefore transition at 1350 K shows the region of beginning of MNMT and the one at 1650 K is a phase transition near the critical temperature.

In spite of the fact that the linear correlation coefficient squared ($R^2$) does not represent a physical quantity, the pattern of variation of $R^2$ of the isotherm Eq. (2) with temperature could identify these transitions too. (See Figure 7.) The variation of this coefficient significantly indicates transitions at 600 K, 1000 K, 1350 K, and 1600 K. To our knowledge the method presented in this study is for the first time and may be treaded as a unique method, which uses equilibrium interaction potential to identify several phase transitions in liquid cesium metal.

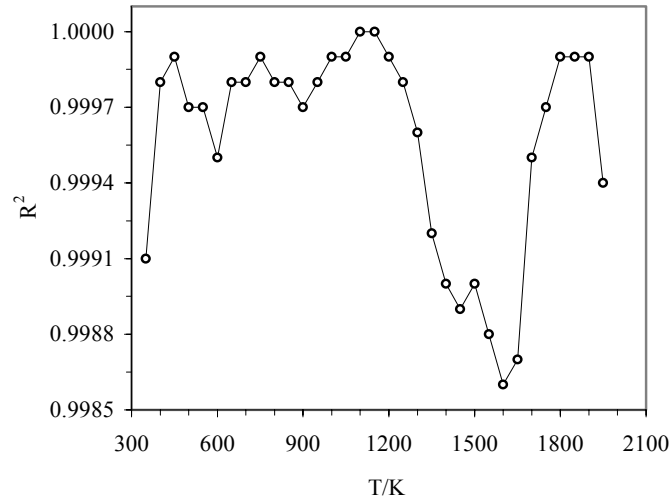

**Figure 7**. Plot of linear correlation coefficient squared ($R^2$) of isotherms Eq. (2) for liquid cesium versus temperature

## 4. Conclusions

The new equation of state based on the semi-empirical (6-3)-potential function has been used to identify phase transitions of liquid cesium. The molecular potential parameters $\varepsilon$ and $r_m$ show second order phase transitions at about 575 K, 800 K, 1000 K, 1350 K, and 1650 K. For transitions at 575 K ($\approx$590 K), 1360 K and 1650 K, experimental evidences are exist in literature, but the transitions at 800 K and 1000 K must be verified experimentally. It has shown that the equilibrium thermodynamic properties can be used as a tool to investigate phase transitions. These properties are alongside with the available (transport property) viscosity, electrical resistance, heat capacity, and structure factor data characteristics of phase transition at 575 K.

The value of $\varepsilon$ at 550-650 K, drops by 5.5%, whereas in the 1350 K region drops by 7.8%. In addition, the variation of $\varepsilon$ with temperature in the former is sharp but in the later one is wide, designating a structureless transition characteristic of a metal non-metal transition. Consequently, during the phase transition, the nature of internal conversion at 575 K is different from that at 1350 K.

**Acknowledgements**

The authors are indebted to the research council of Shiraz University for supporting his study grant number 79-SC-1369-C123

**References:**

[1]   R. Winter, F. Hensel J. Phys. Chem. 92 (1988) 7171-7174.

[2]   F. Hensel H. Uchtmann Annu. Rev. Phys. Chem. 40 (1989) 61-68.

[3]   F. Hensel Phil. Trans. Soc. Lond. A 356 (1998) 97-117.

[4]   L.A. Blagnrovov, S.A. Orlov, S.N. Skovord'ko, V.A. Alekseev High Temp. 38 (2000) 566-572.

[5]   L.A. Blagnrovov, S.N. Skovord'ko, A.S. Krylov, L.A. Orlov, V.A. Alekseev, E.E. Shpilrain J. Non-cryst. Solids **277** (2000) 182-187.

[6]   Yu.I. Sharykin, V.P. Glazkov, S.N. Skovord'ko Doki. Acad. Sci. USSR 244 (1979) 78-82.

[7]   A.Yu. Astapkovich, E.M. Iolin, V.O. Nikolaev, S.N. Skovord'ko, Doki. Acad. Sci. USSR 263 (1982) 73-75.

[8]   M.H. Ghatee, H. Shams-Abadi J. Phys. Chem. B 105 (2001) 702-710.

[9]   M.H. Ghatee, M. Bahadori J. Phys. Chem. 105 (2001) 11256-11263.

[10]   S. Munejiri, F. Shimojo, K. Hoshino, M. Watabe J. Phys.: Condens. Matter 9 (1997) 3303-3312.

[11]   V.F. Kozhevnikov. S.P. Naurzakov, A.P. Senchankov J. Moscow Phys. Soc. 1 (1991) 171-197.

[12]   M.H. Ghatee, M. Sanchooli Fluid Phase Equilibria 214 (2003) 197-207.

[13]   M.H. Ghatee, M. Bahadori J. Phys. Chem. B 108 (2004) 4141-4146.

[14]   H. Weickenmeier, U. Diemar, M. Wahl, M. Raab, W. Demtroder, W. Muller J. Chem. Phys. 82 (1985) 5354-5363; M. Krauss, W. Stevenes J. Chem. Phys. 97 (1990) 4236-4240.

[15]   R. Winter, F. Hensel, T. Bodensteiner, W. Glaser Ber. Bunsenges. Phys. Chem. 91 (1987) 1327-1330.

[16]  R. Winter, F. Hensel Phys. Chem. Liq. **2**0 (1989) 1-15.

[17]  N.H. March Phys. Chem. Liq 20 (1989) 241-245; J. Math. Chem. 4 (1990) 271-287.

[18]  N.H. March Angel Rubio Phys. Rev. B 56 (1997) 13865-13871.

[19]  R.W. Ohse, *Handbook of Thermodynamic and Transport Properties of Alkali Metals*, Oxford, London, 1985, p 485; N.B. Vargaftik, E.B. Gelman, V.F. Kozhevnikov, S.P. Naursako *Int. J. Thermophys*. 11 (1990) 467-476.

[20]  G. Parsafar, F. Kermanpour, B. Najafi J. Phys. Chem. 103 (1999) 7287-7292.

[21]  H. Renkert, F. Hensel, E.U. Frank Ber. Bunsenges. Phys. Chem. 75 (1971) 507-12.

[22]  H.E. Pfeifer, W. Freyland, F. Hensel Ber. Bunsenges. Phys. Chem. 80 (1976) 716-18; 83 (1979) 204-211.

[23] G. Franz, W. Freyland, F. Hensel J. Phys (Paris) Colloq 41 (1980) C8-74-76.

[24]  F. Noll, W.-C. Pilgrim, R. Winter Z. Phys. Chem. Neue Flog 156 (1988) 303-307.

**List of symbols**

Z    Compressibility

P    Pressure

V    molar volume

T    Absolute temperature

R    Gas constant

u    Intermolecular pair potential function

r    Interparticle distance

N    Avogadro's number

k    Boltzman constant

C    Intercept of linear isotherms

B    Slope of linear isotherm

$R^2$    Linear correlation coefficient squared

$K_{bcc}$    Cell constant

*a*    Proportionality constant

**Greek Symbols**

σ    hard-sphere diameter

ε    potential well-depth

ρ    molar liquid density

**Subscript**

m    potential minimum